\documentclass[12pt, a4paper]{article}

\usepackage{amsmath,amssymb,bm,epsfig,afterpage}
\usepackage{cite}
\usepackage{here}
\usepackage{color}
\usepackage{colortbl}
\usepackage{xcolor}
\usepackage{tabularx}
\usepackage{subcaption}
\usepackage{bbm}
\usepackage{graphicx}
\usepackage{listings}
\usepackage{longtable}
\usepackage[utf8]{inputenc}
\usepackage{cancel}
\usepackage{slashed}

\makeatletter
    
    \@addtoreset{equation}{section}
  \makeatother

\allowdisplaybreaks[3]

\newcommand{\Zp}{{Z^\prime}}
\newcommand{\gp}{{g^\prime}}

\newcommand{\MeV}{\mathrm{MeV}}
\newcommand{\GeV}{\mathrm{GeV}}
\newcommand{\TeV}{\mathrm{TeV}}

\newcommand{\eps}{\epsilon}

\newcommand{\Mcal}{\mathcal{M}}
\newcommand{\Lcal}{\mathcal{L}}
\newcommand{\Acal}{\mathcal{A}}

\newcommand{\tU}{\widetilde{U}}

\newcommand{\SM}{\mathrm{SM}}

\newcommand{\ol}[1]{\overline{#1}}
\newcommand{\wt}[1]{\widetilde{#1}}
\newcommand{\vev}[1]{\langle{#1}\rangle}
\newcommand{\abs}[1]{\left|{#1}\right|}
\newcommand{\order}[1]{\mathcal{O}\left({#1}\right)}

\newcommand{\id}[1]{\mathbf{1}_{{#1}} }

{

\newcommand{\eff}{\mathrm{eff}}

\newcommand{\nul}{\nu_\ell} 
\newcommand{\nup}{{\nu^\prime}}
\newcommand{\Dp}{{D^\prime}}
\newcommand{\Lp}{{L^\prime}}
\newcommand{\Wp}{{W^\prime}}
\newcommand{\tnu}{{\tilde{\nu}}}
\newcommand{\tn}{{\tilde{n}}}
\newcommand{\tN}{{\tilde{N}}}
\newcommand{\np}{{n^\prime}}

\newcommand{\tnup}{{\widetilde{\nu}^\prime}}
\newcommand{\tB}{\widetilde{B}}
\newcommand{\tW}{\widetilde{W}}

\newcommand{\tu}{\tilde{u}}

\newcommand{\eV}{\mathrm{eV}}
\newcommand{\MNS}{\mathrm{PMNS}}
\newcommand{\ndbd}{{0\nu\beta\beta}}

\newcommand{\vHp}{v_{H^\prime}}
\newcommand{\keV}{\mathrm{keV}}

\usepackage[colorlinks=true, linkcolor=black, citecolor=black,
urlcolor=black]{hyperref}

\setcounter{footnote}{0}

\newif\iffigsame
\figsamefalse

\begin{document}

\begin{titlepage}

\begin{flushright}
 {\tt
CTPU-PTC-22-26
}
\end{flushright}

\vspace{1.2cm}
\begin{center}
{\Large
{\bf
A Right-handed Neutrino Portal to the Hidden sector : \\ Active Neutrinos and their Twins in an F-theory model
}
}
\vskip 2cm
Junichiro Kawamura$^{a,b}$~\footnote{jkawa@ibs.re.kr}
and
Stuart Raby$^c$~\footnote{raby.1@osu.edu}

\vskip 0.5cm

{\it $^a$
Center for Theoretical Physics of the Universe, Institute for Basic Science (IBS),
Daejeon 34051, Korea
}\\[3pt]

{\it $^b$
Department of Physics, Keio University, Yokohama, 223-8522, Japan
}\\[3pt]

{\it $^c$
Department of Physics, Ohio State University, Columbus, Ohio, 43210, USA}\\[3pt]

\vskip 1.5cm

\begin{abstract}
We analyze the neutrino phenomenology
in an $SU(5)$ F-theory model with both a visible sector and a twin hidden sector.
At low energies, the strong and weak scales of the two sectors may differ
but the spectrum of states is described by the MSSM (MSSM$^\prime$) in the visible (twin) sectors.
What is special about the model is that there are
right-handed neutrinos which couple to both sectors via Yukawa couplings.
As a result, assuming 3 right-handed neutrinos with a large mass 
much greater than the weak scale, at tree-level the seesaw mechanism results
in 3 massive Majorana neutrinos and 3 massless ones.
The massless neutrinos acquire mass via radiative corrections.
In our analysis, the massless neutrinos are predominantly active neutrinos,
while the massive neutrinos are predominantly sterile neutrinos.
We fit the active neutrino masses and mixing angles
and discuss the phenomenology of the lightest sterile neutrino.
Finally we consider some possible scenarios for cosmology.
\end{abstract}
\end{center}
\end{titlepage}

\tableofcontents
\clearpage

\setcounter{footnote}{0}
\section{Introduction}

In Ref.~\cite{Clemens:2019dts},
a model with $SU(5)\times SU(5)^\prime \times U(1)_X$ gauge symmetry is realized
utilizing Heterotic-F-theory duality and a $4+1$ split of the F-theory spectral
divisor~\cite{Clemens:2019wgx,Clemens:2019mvs}.
The Grand Unified Theory (GUT) surface is invariant under
a $\mathbb{Z}_2$ involution which allows for the gauge symmetry to be broken down
to $G_{\mathrm{SM}}\times G_{\mathrm{SM}^\prime}$,
where $G_\SM$ and $G_{\SM^\prime}$ are the Standard Model (SM) gauge group
and its twin counterpart
with Wilson line symmetry breaking~\footnote{The problem of Wilson line breaking
resulting in massless vector-like exotics, emphasized in Refs. \cite{Donagi:2008ca,Beasley:2008kw}, is resolved in the F-theory
model with a bi-section and the $\mathbb{Z}_2$ involution including a translation by the difference of the two sections.}.
To summarize, the Minimal Supersymmetric Standard Model (MSSM)
is realized after $SU(5)$ breaking by a Wilson line.
There are no vector-like exotics and R-parity
 and a $\mathbb{Z}_4^R$ symmetry arise in this construction.
The $SU(5)^\prime$, broken to $G_{\mathrm{SM}^\prime}$ 
corresponds to the twin sector
whose matter content is the same as that in the MSSM sector,
but with a different value for the GUT scale and the GUT coupling constant~\cite{Clemens:2020grq}.

Models with a twin sector, or sometimes called the mirror sector,
have been described
as the parity solution to the strong CP problem in the literature~\cite{PhysRevD.41.1286,Barr:1991qx,Gu:2011yx,Gu:2012in,Gu:2013nya,Abbas:2017hzw,Abbas:2017vle,Gu:2017mkm,Hall:2018let,Kawamura:2018kut,Dunsky:2019api,Berbig:2022hsm}.
The possibility of light sterile neutrinos from the mirror sector
is discussed in Ref.~\cite{Zhang:2013ama}.
In this case there is the mirror sector of the SM without supersymmetry (SUSY).
There are only two right-handed neutrinos,
so that the light sterile neutrino is explained together
with asymmetric dark matter (DM)
from the right-handed neutrino decays~\cite{An:2009vq}.

In this paper, we study the phenomenology of the neutrinos in the F-theory model.
There are only three generations of right-handed neutrinos
which couple to both the MSSM and twin sectors via Yukawa couplings,
as pointed out in Ref.~\cite{Clemens:2019flx}.
The tiny neutrino masses can be explained by the type-I seesaw mechanism.
Three of the neutrinos are massless at the tree-level
since there are only three generations of right-handed neutrinos.
The masses of these states are generated through loop corrections,
and hence their masses are expected to be
$\sim m_D^2/(16\pi^2 M)$,
where $m_D$ is the Dirac neutrino mass which may be at the electroweak (EW) scale
and $M$ is the Majorana mass.
The masses of the other three states are $\sim m_{D^\prime}^2/M$,
where $m_{D^\prime}$ is the Dirac mass for the twin neutrinos
which may be at the twin EW scale.
Thus there are three sterile neutrinos
whose mass are also tiny due to the type-I seesaw mechanism at the tree-level.
We shall study the phenomenology of the active neutrinos
and their mixing with the sterile neutrinos.

The rest of this paper is organized as follows.
In Sec.~\ref{sec-model},
the neutrino sector of the model is introduced,
and then the dominant radiative corrections are calculated.
The neutrino phenomenology is discussed in Sec.~\ref{sec-pheno}.
Section~\ref{sec-concl} concludes.
The results with the soft parameters using mirage mediation
are shown in App.~\ref{sec-mirage}.

\section{Model}
\label{sec-model}

\subsection{Neutrino masses and interactions}
\label{eq-intNu}

The superpotential involving the Majorana neutrinos $N_i$, $i=1,2,3$,
is given by
\begin{align}
 W = \frac{1}{2} N_i M_{ij} N_j
     + N_i Y_{ij} \ell_j H_u + N_i Y_{ij}^\prime \ell^\prime_j H^\prime_u,
\end{align}
where $\ell_i$ and $H_u$ ($\ell_i^\prime$ and $H_u^\prime$)
are the MSSM (twin) lepton and up-type Higgs doublets~\footnote{
Throughout this paper,
primed symbols represent that these are twin counterparts of ones in the MSSM.
}.
Throughout this paper,
we assume that the Majorana neutrino mass
is at the intermediate scale $\sim 10^{12}~\GeV$
much larger than the SUSY breaking scale.

The neutrino masses at the tree-level are given by
\begin{align}
 - \Lcal_{\nu} = \frac{1}{2} n \Mcal_n n + \frac{1}{2} n^\dag \Mcal_n^* n^\dag,
\end{align}
 where
\begin{align}
 n :=
\begin{pmatrix}
 \nu \\ \nup \\ N
\end{pmatrix},
\quad
\Mcal_n =
\begin{pmatrix}
 0 & 0 & m_D^T \\
 0 & 0 & m_\Dp^T \\
m_D & m_\Dp & M
\end{pmatrix}.
\end{align}
Here, $\nu$, $\nup$ and $N$ are the Weyl fermions in the chiral superfields
of the neutral components of $\ell$, $\ell^\prime$ and $N$, respectively.
The elements in $\Mcal_n$ are $3\times 3$ matrices
and the indices $i,j$ are omitted.
The Dirac mass matrices are defined as $m_D := Y v_u$
and $m_{D^\prime} := Y^\prime v_u^\prime$.
The Higgs doublets are expanded around their vacuum expectation values (VEVs)
as
\begin{align}
 H_u =
\begin{pmatrix}
 0 \\ v_u
\end{pmatrix}
+ \frac{1}{\sqrt{2}}
\begin{pmatrix}
 H_u^+ \\ h_u^0 + ia_u
\end{pmatrix},
\quad
 H_d =
\begin{pmatrix}
 v_d \\ 0
\end{pmatrix}
+ \frac{1}{\sqrt{2}}
\begin{pmatrix}
 h_d + ia_d \\ H_d^-
\end{pmatrix}.
\end{align}
The twin Higgs doublets are expanded in the same way.
In general, the $3\times 6$ block $(m_D, m_\Dp)$ is decomposed as
$(m_D, m_\Dp) = u_3^\dag (0, d_3) u_6$,
where $u_3$ ($u_6$) is a $3\times 3$ ($6\times 6$) unitary matrix
and  $d_3$ is a diagonal matrix, by the singular value decomposition,
and thus there are three massless states at the tree-level.

We define the neutrino mass basis as
\begin{align}
n = U \hat{n},
\quad
U^T \Mcal_n U = D_n,
\end{align}
where $D_n$ is a diagonal matrix with positive entries
and $U$ is the $9\times 9$ unitary matrix.
We denote the $3\times 9$ blocks of the unitary matrix as
\begin{align}
 U =
\begin{pmatrix}
 A_L \\ A_\Lp \\ A_R^*
\end{pmatrix},
\end{align}
then the unitarity of $U$ indicates
\begin{align}
\label{eq-ALRone}
 A_L^\dag A_L + A_\Lp^\dag A_\Lp + A_R^T A_R^* = \id{9},
\quad
A_L A_L^\dag = A_\Lp A_\Lp^\dag = A_R A_R^\dag = \id{3},
\quad
A_L A_\Lp^\dag = A_{L_1} A_R^T = 0,
\end{align}
and the diagonalization formula indicates
\begin{align}
\label{eq-ALRtwo}
 A_{L_1} D_n A_{L_2}^T   = 0,
\quad
A_R D_n A_L^\dag = m_D,
\quad
A_R D_n A_\Lp^\dag = m_\Dp,
\quad
 A_R D_n A_R^T = M.
\end{align}
Here, $\{L_1,L_2\} = \{L, L^\prime\}$.

For convenience, we define the 4-component Majorana neutrino as
\begin{align}
 \Psi_I =
\begin{pmatrix}
 \hat{n}_I \\ \hat{n}^\dag_I
\end{pmatrix},
\quad
I = 1,2,3,\cdots,9.
\end{align}
In the mass basis, the $Z$ boson coupling is given by
\begin{align}
 \Lcal_Z = \frac{g}{4c_W} Z_\mu \ol{\Psi} \gamma^\mu
  \left(A_L^\dag A_L P_L - A_L^T A_L^* P_R  \right) \Psi,
\end{align}
where $g$ is the $SU(2)_L$ gauge coupling constant
and $c_W$ is cosine of the weak mixing angle.
Those for the twin $\Zp$ boson is obtained
by replacing $g\to \gp$, $c_W \to c_\Wp$, $Z\to \Zp$, $L\to \Lp$.
The Yukawa couplings to the CP-even Higgs bosons are given by
\begin{align}
 -\Lcal_H = \frac{g}{4c_W m_Z s_\beta}\left(c_\alpha h + s_\alpha H\right)
 \ol{\Psi} \left[
 \left(A_R^\dag m_D A_L + A_L^T m_D^T A_R^* \right) P_L
   + \left( A_R^T m_D^* A_L^* + A_L^\dag m_D^\dag A_R \right) P_R
  \right] \Psi,
\end{align}
and those to the CP-odd ones are
\begin{align}
 -\Lcal_A = \frac{ig}{4c_W m_Z s_\beta}\left(s_\beta a_Z + c_\beta A\right)
 \ol{\Psi} \left[
 \left(A_R^\dag m_D A_L + A_L^T m_D^T A_R^* \right) P_L
   - \left( A_R^T m_D^* A_L^* + A_L^\dag m_D^\dag A_R \right) P_R
  \right] \Psi,
\end{align}
where $a_Z$ is the Nambu-Goldstone mode absorbed by the $Z$ boson.
Here, the mixing angles of the Higgs bosons are defined as
 \begin{align}
\begin{pmatrix}
 h_u \\ h_d
\end{pmatrix}
=
\begin{pmatrix}
 c_\alpha & s_\alpha \\ -s_\alpha & c_\alpha
\end{pmatrix}
\begin{pmatrix}
h \\ H
\end{pmatrix},
\quad
\begin{pmatrix}
 a_u \\ a_d
\end{pmatrix}
=
\begin{pmatrix}
 s_\beta & c_\beta \\ -c_\beta & s_\beta
\end{pmatrix}
\begin{pmatrix}
a_Z \\ A
\end{pmatrix}.
\end{align}
The angle $\beta$ is defined as $\tan\beta := v_u/v_d$,
and $c_\theta := \cos\theta$ ($s_\theta := \sin\theta$) for $\theta = \alpha, \beta$.

\subsection{Sneutrino masses and interactions}
\label{eq-intsNu}

The soft SUSY breaking terms relevant to the neutrino masses are given by
\begin{align}
 -\Lcal_{\mathrm{soft}} =
 \frac{1}{2} M_1 \tB \tB + \frac{1}{2} M_2 \tW \tW
   + \tnu^\dag m_L^2 \tnu + \tN^\dag m_R^2 \tN
+ \left(v_u \tN^T A_n \tnu + \frac{1}{2} \tN^T b_n \tN + h.c. \right),
\end{align}
where $\tB$, $\tW$, $\tnu$ and $\tN$
are the bino, wino, left-handed and right-handed sneutrino, respectively.
In addition, there are the twin sector counterparts, except for the $b_n$ term which is common to both sectors.
The sneutrino mass squared matrix is given by
\begin{align}
 -\Lcal_{\tnu} = \frac{1}{2} \phi_n^\dag \Mcal_\tn^2 \phi_n,
\quad
 \phi_n :=
\begin{pmatrix}
 \tn \\ \tn^*
\end{pmatrix},
\quad
\tn :=
\begin{pmatrix}
 \tnu \\ \tnup \\ \tN
\end{pmatrix},
\end{align}
where
\begin{align}
\Mcal_\tn^2 :=
\begin{pmatrix}
 H & B^* \\ B & H^*
 \end{pmatrix},
\end{align}
with
\begin{align}
H := &\
 \begin{pmatrix}
 m_D^\dag m_D + m_L^2 + \Delta_D & m_D^\dag m_\Dp & m_D^\dag M \\
 m_\Dp^\dag m_D & m_\Dp^\dag m_\Dp + m_\Lp^2 + \Delta_\Dp  & m_\Dp^\dag M  \\
 M^\dag m_D  & M^\dag m_\Dp & M^\dag M + m_D m_D^\dag + m_\Dp m_\Dp^\dag + m_R^2
 \end{pmatrix},
\\
B :=&\
\begin{pmatrix}
0 & 0 & X_n^T \\
0 & 0 & X_\np^T \\
X_n & X_\np & b_n
\end{pmatrix}.
\end{align}
Here, $\Delta_D := (c_{2\beta} m_Z/2)\; \id{3}$
and $X_n := v_u A_n + \mu^* m_D \cot\beta$,
where $\mu$ is the $\mu$ parameter in the superpotential, $W\supset \mu H_u H_d$.
Note that $H^\dag = H$ and $B^T = B$.

This matrix can be diagonalized
by a specific form of the unitary matrix~\cite{Dedes:2007ef},
\begin{align}
\label{eq-tUXY}
\tU^\dag \Mcal_\tn^2 \tU =  D_\tn^2,
\quad
 \tU =
\begin{pmatrix}
 X & i Y \\ X^* & -i Y^*
\end{pmatrix},
\end{align}
where $D_\tn^2$ is the diagonal matrix with positive entries.
From the unitarity of $\tU$, $X$ and $Y$ satisfy
\begin{align}
\label{eq-XY}
XX^\dag + Y Y^\dag  = \id{9},
\quad
X X^T = Y Y^T,
\quad
\mathrm{Re}\left(X^\dag X \right) = \mathrm{Re}\left(Y^\dag Y\right) = \frac{1}{2}\id{9},
\quad
\mathrm{Im}\left(X^\dag Y\right) = 0.
\end{align}
The explicit form after multiplying $\tU$ is given by
\begin{align}
\label{eq-UsnU}
 \tU^\dag \Mcal_\tn^2 \tU =
2
\begin{pmatrix}
 \mathrm{Re}\left(X^\dag H X + X^T B X \right)
& - \mathrm{Im}\left(X^\dag H Y + X^T B Y \right)   \\
 \mathrm{Im}\left(Y^\dag H X - Y^T B X \right)
&
 \mathrm{Re}\left(Y^\dag H Y - Y^T B Y \right)
\end{pmatrix}.
\end{align}
The mass basis is defined as
\begin{align}
 \Phi =
\begin{pmatrix}
 \phi_X \\ \phi_Y
\end{pmatrix}
=
\tU^\dag  \phi_n
=
\begin{pmatrix}
 X^\dag  \tn + X^T \tn^* \\
 -i(Y^\dag \tn - Y^T \tn^*)
\end{pmatrix}.
\end{align}
Note that $\Phi^* = \Phi$, so $\phi_X$ and $\phi_Y$ are real scalar fields.
We decompose $X$ and $Y$ as
\begin{align}
 X = \frac{1}{\sqrt{2}}
\begin{pmatrix}
 X_L \\ X_\Lp \\ X_R^*
\end{pmatrix},
 \quad
 Y = \frac{1}{\sqrt{2}}
\begin{pmatrix}
 Y_L \\ Y_\Lp \\ Y_R^*
\end{pmatrix},
\end{align}
then
\begin{align}
\label{eq-idXY}
&
\frac{1}{2} \left( X_{L_1} X_{L_1}^\dag  + Y_{L_1} Y_{L_1}^\dag \right)
=
\frac{1}{2} \left( X_R X_R^\dag  + Y_R Y_R^\dag \right)
= \id{3},
\quad
X_L X_\Lp^\dag + Y_L Y_\Lp^\dag
=
X_{L_1} X_R^T + Y_{L_1} Y_R^T
= 0, \notag \\
&
X_{L_1} X_{L_2}^T = Y_{L_1} Y_{L_2}^T,
\quad
X_{L_1} X_{R}^\dag = Y_{L_1} Y_R^\dag,
\quad
X_R X_R^T = Y_R Y_R^T,
\\ \notag
&
\mathrm{Re}\left(X_L^\dag X_L + X_\Lp^\dag X_\Lp + X_R^T X_R^*  \right)
 =
\mathrm{Re}\left(Y_L^\dag Y_L + Y_\Lp^\dag Y_\Lp + Y_R^T Y_R^*  \right)
= \id{9},  \\ \notag
&
 \mathrm{Im}\left(X_L^\dag Y_L + X_\Lp^\dag Y_\Lp + X_R^T Y_R^*  \right) = 0,
\end{align}
where $\{L_1, L_2\} = \{L, \Lp\}$.
The scalar fields in the gauge basis are related to $\phi_{X,Y}$ as
\begin{align}
 \tnu = \frac{1}{\sqrt{2}}  \left(X_L \phi_X + i Y_L \phi_Y\right),
\quad
 \tnu^\prime = \frac{1}{\sqrt{2}}  \left(X_\Lp \phi_X + i Y_\Lp \phi_Y\right),
\quad
 \tN = \frac{1}{\sqrt{2}}  \left(X_R^* \phi_X + i Y_R^* \phi_Y\right).
\end{align}

The neutrino-sneutrino-neutralino interactions in the mass basis are given by
\begin{align}
 - \Lcal_{\nu\tnu\chi}
 =  \phi^T_X \ol{\psi}_\chi \left( h_{X_L} P_L + h_{X_R} P_R \right) \Psi
    +  \phi^T_Y \ol{\psi}_\chi \left( h_{Y_L} P_L + h_{Y_R} P_R \right) \Psi,
\end{align}
where
\begin{align}
\label{eq-hXYLR}
 h_{X_L} :=&\ \frac{1}{\sqrt{2}v_u} \left[
    V_{H_u}^T  \left(X_R^\dag m_D A_L + X_L^T m_D^T A_R^* \right)
    + s_\beta m_W  V_G^T X_L^\dag A_L \right]
   , \\ \notag
 h_{X_R} :=&\ \frac{1}{\sqrt{2}v_u}
    \left[ V_{H_u}^\dag \left(X_L^\dag m_D^\dag A_R + X_R^T m_D^* A_L^* \right)
      + s_\beta m_W V_G^\dag X_L^T A_L^* \right],  \\ \notag
 h_{Y_L} :=&\ \frac{i}{\sqrt{2}v_u} \left[
    V_{H_u}^T  \left(Y_R^\dag m_D A_L + Y_L^T m_D^T A_R^* \right)
    - s_\beta m_W  V_G^T Y_L^\dag A_L \right]
   , \\ \notag
 h_{Y_R} :=&\ \frac{-i}{\sqrt{2}v_u} \left[
     V_{H_u}^\dag  \left(Y_L^\dag m_D^\dag A_R + Y_R^T m_D^* A_L^* \right)
    - s_\beta m_W  V_G^\dag Y_L^T A_L^* \right].
\end{align}
Here, the neutralinos are defined as
\begin{align}
\begin{pmatrix}
 \tilde{B} \\ \tilde{W}^0 \\ \tilde{H}_d^0 \\ \tilde{H}_u^0
\end{pmatrix}
= V_\chi \chi
=:
\begin{pmatrix}
 V_B \\ V_W \\ V_{H_d} \\ V_{H_u}
\end{pmatrix}
\chi
,
\quad
 V_\chi^T \Mcal_\chi V_\chi  = D_\chi
:= \mathrm{diag}\left(m_{\chi_1},m_{\chi_2},m_{\chi_3},m_{\chi_4} \right),
\end{align}
with the neutralino mass matrix given by
\begin{align}
  \Mcal_\chi
=
\begin{pmatrix}
 M_1 & 0 & -c_\beta s_W m_Z & s_\beta s_W m_Z  \\
 0 & M_2 & c_\beta c_W m_Z  & -s_\beta c_W m_Z \\
-c_\beta s_W m_Z & c_\beta c_W m_Z & 0 & -\mu \\
s_\beta s_W m_Z & -s_\beta c_W m_Z & -\mu & 0
\end{pmatrix}.
\end{align}
The 4-component Majorana neutralino $\psi_\chi$ is defined as
\begin{align}
 \left[\psi_\chi\right]_a =
\begin{pmatrix}
 \chi_a  \\ \chi^\dag_a
\end{pmatrix},
\quad
a=1,2,3,4.
\end{align}
In Eq.~\eqref{eq-hXYLR}, $V_G := V_W - t_W V_B$ with $t_W := s_W/c_W$.

\subsection{Diagonalization for large $M$}

Throughout this paper, we consider the type-I seesaw mechanism,
and hence $M$ is assumed to be
much larger than the SUSY breaking and the (twin) EW scale.
We here derive the diagonalization of the neutrino and sneutrino mass matrices
at the leading order in $M^{-1}$.

The neutrino mass matrix is approximately diagonalized by
\begin{align}
 U = U_1 U_2  :=
\begin{pmatrix}
 \id{3} & 0 & m_D^\dag M_*^{-1} \\
 0 & \id{3} &  m_\Dp^\dag M_*^{-1} \\
 -M^{-1}m_D & - M^{-1} m_\Dp  & \id{3}
\end{pmatrix}
\begin{pmatrix}
u_\ell & 0 \\ 0 & u_N
\end{pmatrix}
+ \order{M^{-2}},
\end{align}
such that
\begin{align}
U^T \Mcal_n U
\simeq
\begin{pmatrix}
d_{n_\ell} & 0 \\ 0 & d_N
\end{pmatrix}.
\end{align}
Here, $M_*^{-1} := (M^*)^{-1}$.
The $6\times6$ ($3\times 3$) unitary matrix $u_\ell$ ($u_N$) satisfies,
\begin{align}
 u_\ell^T
 \begin{pmatrix}
  - m_D^T M^{-1} m_D &   - m_D^T M^{-1} m_\Dp \\
  - m_\Dp^T M^{-1} m_D &   - m_\Dp^T M^{-1} m_\Dp \\
 \end{pmatrix}
u_\ell
= d_{n_\ell}
, \quad
u_N^T M u_N = d_N,
\end{align}
where $d_{n_\ell}$ and $d_N$ are diagonal matrices.

Next, we shall diagonalize the sneutrino mass matrix.
The matrices $X$ and $Y$ in the unitary matrix $\tU$, see Eq.~\eqref{eq-tUXY},
are given by~\cite{Dedes:2007ef}
\begin{align}
\label{eq-XYapp}
X = \frac{1}{\sqrt{2}} \tu
\begin{pmatrix}
 \id{6} + R & 0 \\  0 & \id{3}
\end{pmatrix},
\quad
Y = \frac{1}{\sqrt{2}} \tu
\begin{pmatrix}
 \id{6} - R & 0 \\  0 & \id{3}
\end{pmatrix},
\end{align}
where
\begin{align}
 \tu \simeq
U_1
\begin{pmatrix}
 u_L & 0 & 0 \\
 0 & u_{\Lp} & 0  \\
 0 & 0 & u_R
\end{pmatrix}
,
\quad
\left[R\right]_{ii} = 0,
\quad
\left[R\right]_{ij} = \frac{\left[\Delta\right]_{ij}^*}{d_{L_j}^2-d_{L_i}^2}
\quad (i\ne j).
\end{align}
Here, $U_1$ is the same as for the neutrinos.
The $3\times 3$ unitary matrices $u_L$, $u_\Lp$ and $u_R$ satisfy
\begin{align}
 u_L^\dag \left(m_L^2 + \Delta_D \right) u_L
  =&\ \mathrm{diag}\left(d_{L_1}^2, d_{L_2}^2, d_{L_3}^2 \right),
\quad
 u_\Lp^\dag \left(m_\Lp^2 + \Delta_\Dp \right) u_\Lp
  = \mathrm{diag}\left(d_{L_4}^2, d_{L_5}^2, d_{L_6}^2 \right),  \notag \\
 u_R^\dag \left( M^\dag M + m_R^2 \right) u_R
  =&\  \mathrm{diag}\left(d_{R_1}^2, d_{R_2}^2, d_{R_3}^2 \right).
\end{align}
The matrix $\Delta$ is defined as
\begin{align}
&   \Delta
:=
\begin{pmatrix}
 u_L^T & 0 \\0 &  u_\Lp^T
\end{pmatrix}
\left(
\begin{matrix}
 - X^T_n M^{-1} m_D - m_D^T M^{-1} X_n + m_D^T M^{-1} b_n M^{-1} m_D  \\
- X^T_\np M^{-1} m_D - m_\Dp^T M^{-1} X_n + m_\Dp^T M^{-1} b_n M^{-1} m_D
 \end{matrix}
\right.  \\ \notag
&\ \hspace{4.5cm}
\left.
 \begin{matrix}
- X^T_n M^{-1} m_\Dp - m_D^T M^{-1} X_\np + m_D^T M^{-1} b_n M^{-1} m_\Dp   \\
-X^T_\np M^{-1} m_\Dp - m_\Dp^T M^{-1} X_\np + m_\Dp^T M^{-1} b_n M^{-1} m_\Dp \\
\end{matrix}
\right)
\begin{pmatrix}
 u_L & 0 \\0 &  u_\Lp
\end{pmatrix},
\end{align}
where $b_n \sim \order{m_\mathrm{SUSY} M}$ is assumed.
Here the phase convention of $u_L$ and $u_\Lp$ are chosen
such that  the diagonal elements of $\Delta$ are positive.
Note that $X,Y$ in Eq.~\eqref{eq-XYapp} satisfy Eq.~\eqref{eq-XY}
for the unitarity of $\tU$ up to $\order{M^{-2}}$.
Inserting Eq.~\eqref{eq-XYapp} into Eq.~\eqref{eq-UsnU},
\begin{align}
\tU^\dag \Mcal_{\tn}^2 \tU =&\
 \begin{pmatrix}
\mathrm{Re}\left( (1+R^\dag) d_L^2 (1+R) +  \Delta  \right) & \cdot
& -\mathrm{Im}\left( (1+R^\dag)d_L^2 (1-R)+\Delta  \right) & \cdot \\
\cdot & d_R^2 &  \cdot & \cdot \\
-\mathrm{Im}\left( (1+R)^\dag d_L^2 (1-R) + \Delta \right) & \cdot &
\mathrm{Re}\left( (1-R^\dag)d_L^2(1-R) - \Delta \right) & \cdot  \\
\cdot & \cdot & \cdot & d_R^2 \\
\end{pmatrix}
+ \order{M^{-2}} \\ \notag
\simeq&\
 \mathrm{Diag}
       \left( d_L^2 + d^2_\Delta, d_R^2, d_L^2 - d^2_\Delta, d_R^2  \right)
=: \mathrm{Diag}\left(D_X^2, D_Y^2\right),
\end{align}
where the entries with $\cdot$ in the first line are $\order{M^0}$
and are irrelevant for the light sneutrinos.
Here, $\mathrm{Diag}$ is a matrix whose diagonal blocks are given by its arguments.
$d_\Delta^2$ is the diagonal matrix whose diagonal elements are the same as $\Delta$.
In our convention with $d_\Delta^2 >0$,
$\phi_X$ is heavier than $\phi_X$ by $\abs{\Delta} = \order{X_n M^{-1} m_D}$.

\subsection{Loop corrections to neutrino masses}

The neutrino masses after integrating out
the right-handed neutrinos are given by
\begin{align}
  -\Lcal_{\mathrm{mass}} =&\ - \frac{1}{2}
                            \left(\nu^T m_D^T + \nu^{\prime T} m_{D^\prime}^T \right)
                            M^{-1} \left(m_D \nu + m_{D^\prime} \nu^\prime \right)  \\ \notag
 =&\ -\frac{1}{2}
\begin{pmatrix}
 \nu \\ \nu^\prime
\end{pmatrix}^T
\begin{pmatrix}
 m_D^T M^{-1} m_D &  m_D^T M^{-1} m_{D^\prime} \\
 m_{D^\prime}^T M^{-1} m_D &  m_{D^\prime}^T M^{-1} m_{D^\prime}
\end{pmatrix}
\begin{pmatrix}
 \nu^T \\ \nu^\prime
\end{pmatrix}
=: \frac{1}{2} n^T_\ell \Mcal_{\nul} n_\ell.
\end{align}
There are three massless states at the tree-level
since the rank of the $6\times6$ mass matrix $\Mcal_{\nul}$ is 3.
We shall assume that $m_{D^\prime} \gg m_D$,
hence the three massless neutrinos are predominantly active neutrinos which acquire their mass
from radiative corrections.

The radiative corrections to the Majorana neutrino masses
in the non-SUSY multi-Higgs doublet model is calculated
in Ref.~\cite{Grimus:2002nk}.
The loop corrections in the MSSM with the right-handed neutrinos
are calculated in Refs.~\cite{Grimus:2002nk,CandiadaSilva:2020hxj}.
We follow the calculation in Ref.~\cite{Grimus:2002nk}
and apply it to the sneutrino loops as well
using the interactions for the 4-component fermions shown in
Sections~\ref{eq-intNu} and \ref{eq-intsNu}.
At the loop-level, the neutrino mass matrix is corrected as~\cite{Grimus:2002nk}
\begin{align}
 \Mcal_{\nu_\ell} \to
 \Mcal_{\nu_\ell} + \delta \Mcal_{\nu_\ell},
\quad
\delta \Mcal_{\nu_\ell}
:=
\begin{pmatrix}
 \delta{m_{LL}} & \delta m_{L\Lp} \\  \delta m_{\Lp L} & \delta m_{\Lp\Lp}
\end{pmatrix}
:= -
\begin{pmatrix}
 A_L^* \Sigma_L A_L^\dag & A_L^* \Sigma_L A_{\Lp}^\dag \\
 A_\Lp^* \Sigma_L A_L^\dag & A_\Lp^* \Sigma_L A_{\Lp}^\dag \\
\end{pmatrix},
\end{align}
where $\Sigma_L$ is the coefficient of $P_L$ in the neutrino self-energy.
Note that the loop diagrams mediated by the MSSM (twin) particles
contribute to only the upper-left (lower-right) block
and the off-diagonal blocks are vanishing,
i.e. $\delta m_{L\Lp} = \delta m_{\Lp L} =0$.
Physically, this is because there is no particle which mediates the MSSM and twin sector
other than the right-handed (s)neutrinos.
Equations~\eqref{eq-ALRone} and~\eqref{eq-ALRtwo} ensure this formally.
The correction from the neutrino-SM boson loop diagrams is given by
\begin{align}
\delta_\nu m_{LL}
 = \frac{g^2}{64\pi^2 m_W^2} m_D^T A_R^* D_n \left(
   \frac{3 \log x_n^Z}{x_n^Z-1} + \sum_{S=h,H,A} c_S \frac{\log x^S_n}{x^S_n-1}
  \right)  A_R^\dag m_D,
\end{align}
where $x^B_n := D_n^2/m_B^2$ for $B=Z,h,H,A$.
Here, $c_h := c_\alpha^2/s_\beta^2$, $c_H := s_\alpha^2/s_\beta^2$
and $c_A := -c_\beta^2/s_\beta^2$.
We used Eqs.~\eqref{eq-ALRone} and~\eqref{eq-ALRtwo} to simplify the result.
This result is consistent with Ref.~\cite{Grimus:2002nk}.
The sneutrino-neutralino contribution is given by
 \begin{align}
\delta_\tnu m_{LL}
 =&\ \sum_{a=1}^4
  \frac{m_{\chi_a}}{32\pi^2 v_u^2} \Biggl[
(V_{H_u}^{a})^2 m_D^T \left(X_R^*g_{\chi_a X}X_R^\dag-Y_R^*g_{\chi_a Y}Y_R^\dag\right)m_D
 \\ \notag
&\ \quad
+ (V_{G}^a)^2 s_\beta^2 m_W^2
 \left( X_L^* g_{\chi_a X} X_L^\dag - Y_L^* g_{\chi_a Y}Y_L^\dag \right) \\ \notag
&\ \quad + s_\beta m_W V_{H_u}^a V_{G}^a
 \left\{
 m_D^T \left(X_R^* g_{\chi_a X} X_L^\dag + Y_R^* g_{\chi_a Y} Y_L^\dag \right)
 + \left(X_L^* g_{\chi_a X}X_R^\dag + Y_L^* g_{\chi_a Y} Y_R^\dag  \right) m_D
  \right\}
\Biggr],
\end{align}
where
\begin{align}
 g_{\chi_a Z} :=
\frac{D_Z^2 }{D_Z^2 - m_{\chi_a}^2}\log\frac{D_Z^2}{m_{\chi_a}^2},
\quad Z=X,Y.
\end{align}
The identities in Eq.~\eqref{eq-idXY} are used to simplify the result.

At the leading order in $M$,
\begin{align}
\delta_\nu m_{LL}
\simeq&\ \frac{g^2}{64\pi^2 c_W^2} m_D^T
   \left[\frac{4+\Delta_H}{2}\left(M^{-1} u_N^* L_{NZ} u_N^T +
   u_N L_{NZ} u_N^\dag M^{-1}
  \right)
- M^{-1} \sum_{S} c_S \frac{m_S^2}{m_Z^2} \log\frac{m_S^2}{m_Z^2}
 \right] m_D, \notag \\
\delta_\tnu m_{LL}
\simeq&\ -\frac{g^2}{32\pi^2 c_W^2} m_D^T
 \left(u_R L_{NZ} u_R^\dag M^{-1}+ M^{-1} u_R^* L_{NZ}u_R^T \right) m_D \\ \notag
&\  + \sum_a \frac{m_{\chi_a}}{16\pi^2 v_u^2}
  \left[ (V_G^a)^2 s_\beta^2 m_W^2 u_L^* \frac{\Delta g_{\chi_a L}}{\Delta m_L^2} u_L^T
 \right. \\  \notag
&\quad\quad \left.
- s_\beta m_W V_G^a V_{H_u}^a
 \left( m_D^T M^{-1} m_D u_L g_{\chi_a L} u_L^\dag
 + u_L^* g_{\chi_a L} u_L^T m_D^T M^{-1}u_L^T
 \right)
  \right],
\end{align}
where $L_{NZ}:= \log(d_N^2/m_Z^2)$ and $\Delta_H := -1 +\sum_{S} c_S ({m_S^2}/{m_Z^2})$.
The loop functions are defined as
\begin{align}
 g_{\chi_a L} :=&\ \frac{d_L^2 }{d_L^2 - m_{\chi_a}^2}
                 \log \frac{d_L^2}{m_{\chi_a}^2},
\quad
 g^\prime_{\chi_a L}:=\frac{d_L^2-m_{\chi_a}^2-m_{\chi_a}^2\log({d_L^2}/{m_{\chi_a}^2})}
                           {\left(d_L^2 - m_{\chi_a}^2\right)^2 }, \notag \\
\left[\frac{\Delta g_{\chi_a L}}{\Delta m_L^2} \right]_{ij}
 := &\ \left[g^\prime_{\chi_a L} \right]_{ii} \delta_{ij}
         + \frac{\left[g_{\chi_a L} \right]_{jj}-\left[g_{\chi_a L} \right]_{ii}}
                {\left[d_L^2\right]_{jj}- \left[d_L^2\right]_{ii}}
                \Delta_{ij} (1-\delta_{ij}),
 \end{align}
where $i,j=1,2,3$ are the indices for the MSSM sneutrinos.
We used $ \sum_{a} V_{H_u}^a V_{G}^a m_{\chi_a} = -{s_\beta m_W}/{c_W^2}$
for the first line of the sneutrino correction.

Altogether, the dominant contribution to $\delta m_{LL}$ is given by
\begin{align}
 \delta m_{LL} \simeq &\
  \frac{g^2}{64\pi^2c_W^2} m_D^T
   \left[\frac{\Delta_H}{2}\left(M^{-1} u_N^* L_{NZ} u_N^T +
   u_N L_{NZ} u_N^\dag M^{-1}
  \right)
 - M^{-1} \sum_{S} c_S \frac{m_S^2}{m_Z^2} \log\frac{m_S^2}{m_Z^2},
 \right] m_D  \notag     \\
&\ + \frac{g^2}{32\pi^2}
    V_G D_\chi u_L^* \frac{\Delta g_{\chi L}}{\Delta m_{L}^2} u_L^\dag V_G^T  \\ \notag
&\ - \frac{s_\beta m_W}{16\pi^2 v_u^2}
 V_G D_\chi
\left(
u_L^* g_{\chi L} u_L^T m_D^T M^{-1} m_D
 + m_D^T M^{-1} m_D u_L g_{\chi L} u_L^\dag  + 2 m_D^T M^{-1} m_D L_{\chi Z}
\right) V_{H_u}.
\end{align}
Since $u_N \simeq u_R$,
the $L_{NZ}$ dependent parts without $\Delta_H$
are canceled between the neutrino and sneutrino contributions.
Hence the $L_{NZ}$ dependence appears only with $\Delta_H$
which vanishes in the SUSY limit,
but $\Delta_H \sim -1 + m_h^2/m_Z^2 \sim 0.88$ is not a small factor
due to the radiative corrections to the SM-like Higgs boson.
The importance of this effect is also addressed in Ref.~\cite{CandiadaSilva:2020hxj}.
As discussed in Refs.~\cite{Grimus:2002nk,Dedes:2007ef},
the dominant contributions to the light neutrino masses
are those directly to the Majorana masses,
and those to the Dirac masses and wave-function renormalization
are sub-dominant.
The loop correction for the twin neutrinos are given by the same form
with formally replacing those in the twin sector.

In Ref.~\cite{Dedes:2007ef},
the authors calculated the contributions from the neutralino-sneutrino loop
which could dominate over the tree-level contribution
by the large soft parameters contained in $\Delta$.
This is not an issue in our
case,  since the tree-level contributions for the active neutrinos vanish to leading order.
They also comment that the SM neutrino-$Z$/Higgs boson loops are a one percent correction.
We include these in our evaluation of the radiative correction to neutrino masses.
In fact, these contributions are important particularly for heavy sneutrinos,
because $g^\prime_{\chi L}\propto d_{L_i}^{-2}$
which appears in the gaugino loop contribution for $d_{L_i} \gg m_{\chi_a}$.

\section{Phenomenology}
\label{sec-pheno}

\subsection{Observables}

We define the diagonalization unitary matrix $U_\ell$
for the light neutrino mass matrix with radiative corrections as
\begin{align}
 U_\ell^T \Mcal_{\nu_\ell} U_\ell = \mathrm{diag}\left(
m_{\nu_1},m_{\nu_2},m_{\nu_3},m_{\nu_4},m_{\nu_5},m_{\nu_6}
\right),
\end{align}
where $\Mcal_{\nu_\ell}$ includes the radiative corrections.
We decompose $U_\ell$ and parametrize the upper $3\times 6$ block of it as
\begin{align}
U_\ell  =:
\begin{pmatrix}
 A_{\ell} \\ A_{\ell^\prime}
\end{pmatrix},
\quad
A_\ell =:
\begin{pmatrix}
 U_{e1} &  U_{e2} &  U_{e3} &  U_{e4} &  U_{e5} &  U_{e6} \\
 U_{\mu1} &  U_{\mu2} &  U_{\mu3} &  U_{\mu4} &  U_{\mu5} &  U_{\mu6} \\
 U_{\tau1} &  U_{\tau2} &  U_{\tau3} &  U_{\tau4} &  U_{\tau5} &  U_{\tau6} \\
\end{pmatrix},
\end{align}
where the left (right) three columns of $A_\ell$ are
for the active (sterile) neutrinos.

The $W$ boson coupling is given by
\begin{align}
 \Lcal_W
= \frac{g}{\sqrt{2}} W_\mu^- \ol{\psi}_e
           \gamma^\mu P_L A_\ell \Psi_\ell + h.c.,
\end{align}
where $\Psi_\ell$ contains the neutrinos in the mass basis
except the heaviest three states with $\order{M}$ masses.
Here, we choose the flavor basis that the charged lepton Yukawa matrix
is positive diagonal in the gauge basis.
We assume that the Pontecorvo-Maki-Nakagawa-Sakata (PMNS) matrix
for the active neutrinos are almost unitary,
so that the angles in the standard parametrization,
\begin{align}
U_\MNS=
\begin{pmatrix}
1 & 0 & 0 \\
0 & c_{23} & s_{23} \\
0 & -s_{23} & c_{23} \\
\end{pmatrix}
\begin{pmatrix}
c_{13} & 0 & s_{13} e^{-i \delta} \\
0 & 1 & 0 \\
-s_{13} e^{i \delta} & 0 & c_{13} \\
\end{pmatrix}
\begin{pmatrix}
 c_{12} & s_{12}& 0 \\
 -s_{12} & c_{12} & 0 \\
0 & 0 & 1
\end{pmatrix},
\end{align}
are related to the elements in $A_\ell$ as
\begin{align}
s_{12} = \frac{\abs{U_{e2}}}{\sqrt{\abs{U_{e1}}^2 + \abs{U_{e2}}^2 }}
\quad
s_{23} = \frac{\abs{U_{\mu 3}}}{\sqrt{\abs{U_{\mu 3}}^2 + \abs{U_{\tau 3}}^2 }}
\quad
 s_{13} = \abs{U_{e3}},
\end{align}
and
\begin{align}
 e^{i\delta}
= \abs{\frac{U_{e2} U_{e3} U_{\mu 3}}{U_{e1}U_{\tau 3}}}
 \left( 1+
\frac{\sqrt{\abs{U_{e1}}^2 + \abs{U_{e2}}^2}\sqrt{\abs{U_{\tau 3}}^2 + \abs{U_{\mu 3}}^2}}
     {\abs{U_{\mu 3}}^2} \frac{U_{\mu 2}^* U_{\mu 3}}{U_{e2}^* U_{e3}}\right).
\end{align}
Here, we consider the fitted data
with the normal ordering (NO)~\cite{deSalas:2017kay,ParticleDataGroup:2022pth}:
\begin{align}
\Delta m_{12}^2 = (7.55 \pm 0.20)\times 10^{-5}~\eV,
\quad
\Delta m_{23}^2 = (2.424 \pm 0.030)\times 10^{-3}~\eV,
\end{align}
\begin{align}
 s_{12}^2= 0.32\pm 0.02,
\quad
 s_{23}^2= 0.547\pm 0.03,
\quad
 s_{13}^2= 0.0216 \pm 0.0083,
\quad
\delta_\mathrm{CP} = 218\pm38~\mathrm{deg},
\end{align}
and with the inverted ordering (IO)~\cite{deSalas:2017kay,ParticleDataGroup:2022pth}:
\begin{align}
\Delta m_{12}^2 = (7.55 \pm 0.20)\times 10^{-5}~\eV,
\quad
\Delta m_{23}^2 = (-2.50 \pm 0.040)\times 10^{-3}~\eV,
\end{align}
\begin{align}
 s_{12}^2= 0.32\pm 0.02,
\quad
 s_{23}^2= 0.5551 \pm 0.03,
\quad
 s_{13}^2= 0.0220 \pm 0.0076,
\quad
\delta_\mathrm{CP} = 281 \pm 27~\mathrm{deg},
\end{align}
where $\Delta m_{ij}^2 := m_{\nu_j}^2 - m_{\nu_i}^2$.
For reference, typical values of the absolute values of the PMNS matrix is
\begin{align}
\label{eq-absPMNS}
 \abs{U_\MNS} \sim
\begin{pmatrix}
 0.82 & 0.55 & 0.15 \\
 0.31 & 0.60 & 0.74 \\
 0.48 & 0.58 & 0.66
\end{pmatrix}.
\end{align}
In our notation, the lightest neutrino is $\nu_1$ ($\nu_3$)
for the NO (IO) case,
but the sterile neutrinos are ordered by their masses,
so the lightest sterile neutrino is always $\nu_4$.

The Majorana neutrinos can induce the neutrino-less double
$\beta$ ($\ndbd$) decay.
The $\ndbd$ decay half-life is given by~\cite{Faessler:2014kka},
\begin{align}
 \left[T^{0\nu}_{1/2} \right]^{-1}
= \Acal\abs{m_p\sum_{i=1}^6 U^2_{ei} \frac{m_{\nu_i}}{\vev{p^2}+m_{\nu_i}^2}}^2,
\end{align}
where the values of $\Acal$ and $\vev{p^2}\simeq (200~\MeV)^2$ are tabulated
in Table.1 of Ref.~\cite{Faessler:2014kka}.
We choose the values which provide the most conservative limits,
i.e. (a) Argonne potential with $g_A=1.00$,
for $\vev{p^2}$ and $\Acal$;
$\sqrt{\vev{p^2}} (^{76}\mathrm{Ge})= 0.159~\GeV$,
$\Acal(^{76}\mathrm{Ge})  = 2.55\times 10^{-10}\;\mathrm{yrs}^{-1}$,
and
$\sqrt{\vev{p^2}} (^{136}\mathrm{Xe})= 0.178~\GeV$,
$\Acal(^{136}\mathrm{Xe})  = 4.41\times 10^{-10}\;\mathrm{yrs}^{-1}$.
The limits for $^{76}$Ge and $^{136}$Xe
are~\cite{KamLAND-Zen:2012mmx,GERDA:2013vls}
\begin{align}
 T_{1/2}^{0\nu}(^{76}\mathrm{Ge})
 \ge 3.0\times 10^{25}~\mathrm{years},
\quad
 T_{1/2}^{0\nu}(^{136}\mathrm{Xe})
 \ge 3.4\times 10^{25}~\mathrm{years},
\end{align}
respectively.

\subsection{Numerical study}

\begin{figure}[t]
 \centering
\includegraphics[height=0.48\hsize]{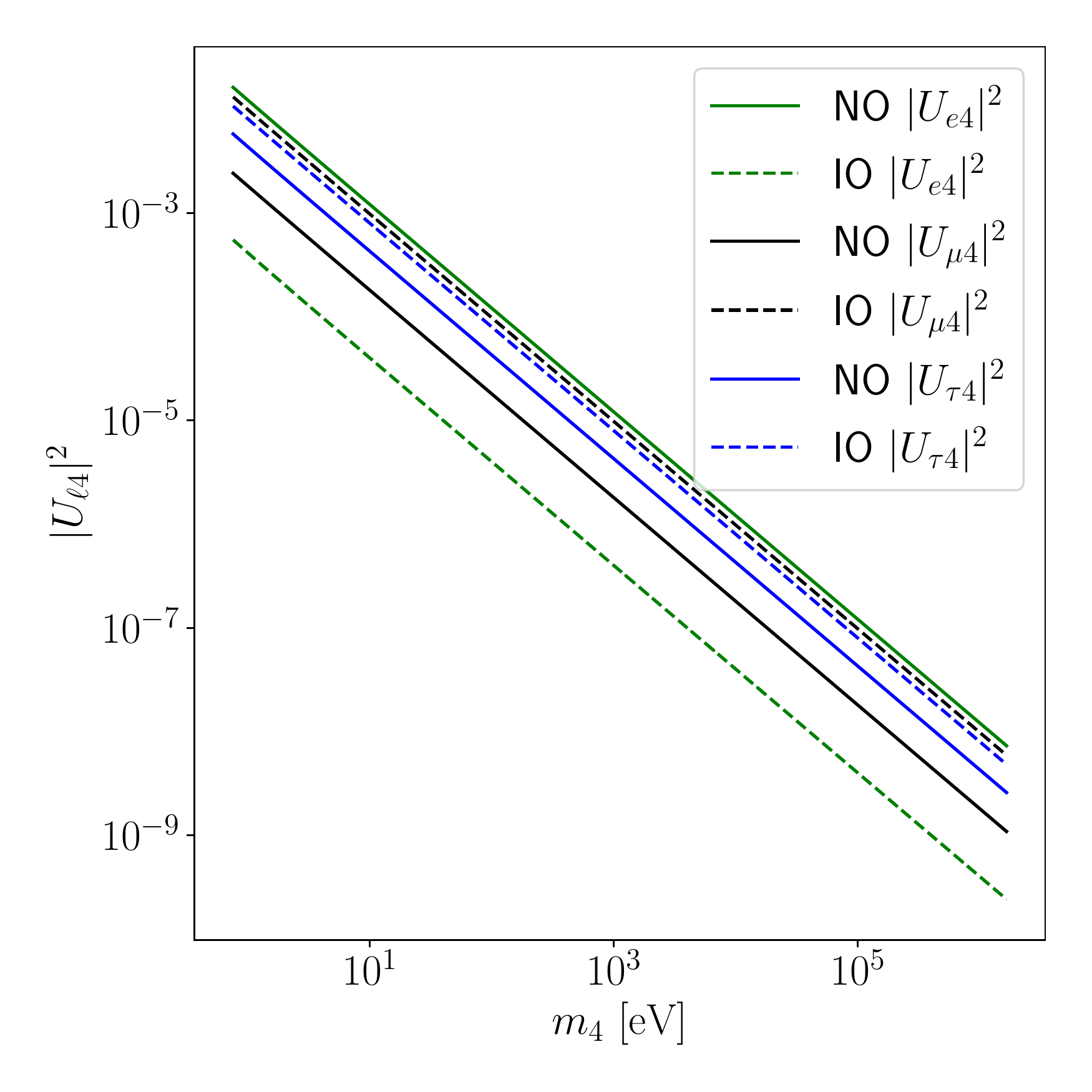}
\includegraphics[height=0.48\hsize]{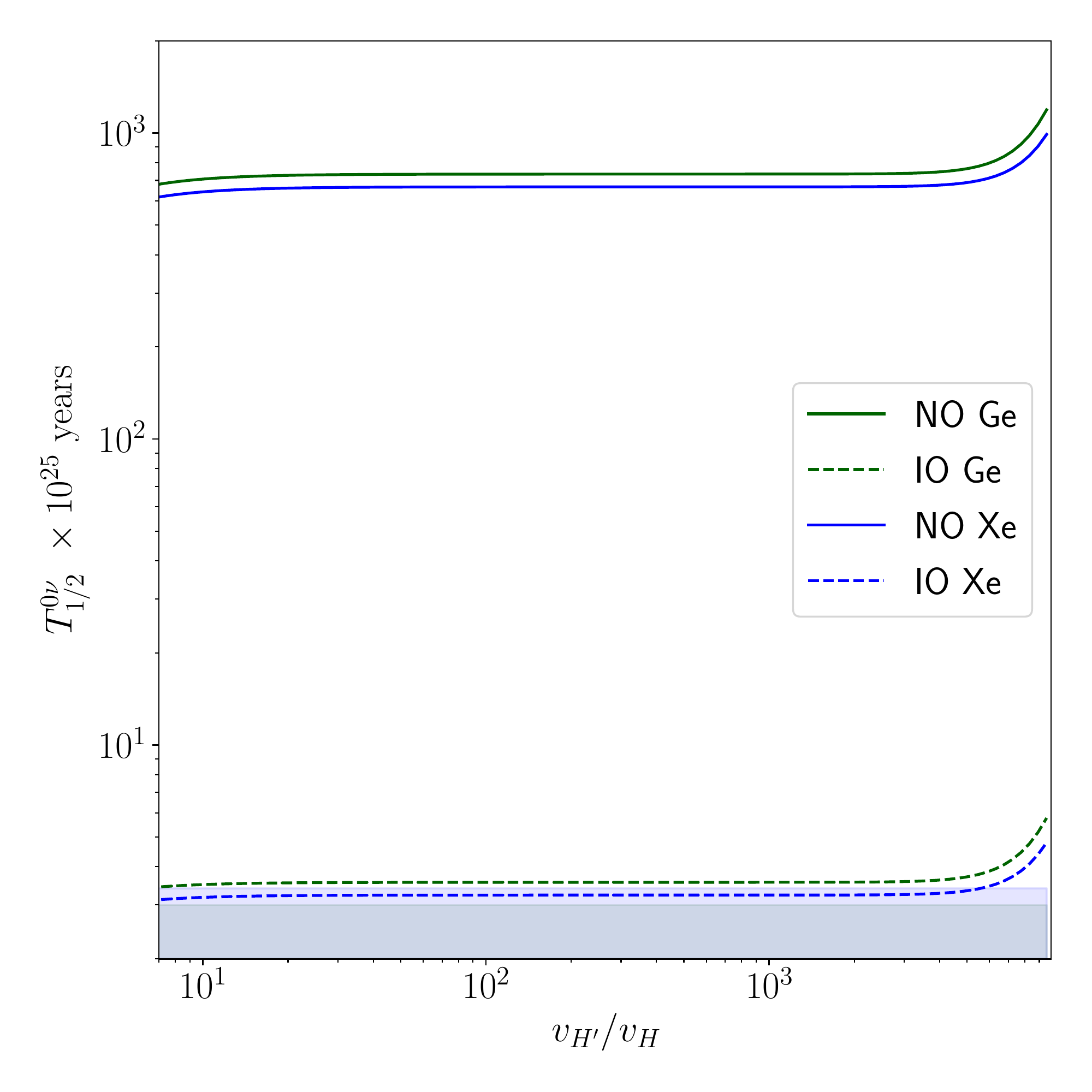}
\caption{\label{fig-nu4CM}
The mass and mixing of the lightest sterile neutrino $\nu_4$ (left)
and the $\ndbd$ decay with varying $\vHp/v_H$ (right).
On the left panel,
the solid (dashed) lines correspond to the NO (IO).
The colors are the charged lepton flavors.
On the right panel,
the green (blue) lines are the values for Ge (Xe).
The colored regions are the lower bounds on the lifetime.
}
\end{figure}

We study the observables in our model with simplified parameter sets.
Without loss of generality, the Majorana mass matrix $M$
and the charged (twin) lepton Yukawa matrix $Y_e$ ($Y_e^\prime$)
are diagonal with positive elements.
For simplicity, we shall assume~\footnote{
It is our future work to study how the Yukawa couplings
are determined from the F-theory construction.
The Yukawa couplings may be different from each other
just like the gauge coupling constants~\cite{Clemens:2020grq},
but we assume here that the flavor structures are common
in the MSSM and twin sector.
}
\begin{align}
Y = Y^\prime,
\quad
v_H < v^\prime_H,
\end{align}
where $v_H := \sqrt{v_u^2 + v_d^2} \sim  174~\GeV$.
Further, we assume that the soft parameters
$m_L^2$, $m_\Lp^2$ and $m_R^2$ are diagonal and real,
and $A_n = A_0 Y$, $A^\prime_n = A^\prime_0 Y^\prime$ and $b_n = B_0 M$,
where $A_0$, $A_0^\prime$ and $B_0$ are scalar constants.
In the Higgs sector, we assume the decoupling limit, $m_A \gg m_h$,
so that $m_H \simeq m_A$ and $c_\alpha^2 \simeq s_\beta^2$.
Finally we introduce the parametrization of the Dirac mass matrices,
\begin{align}
 m_D = M^{\frac{1}{2}} d_n^{\frac{1}{2}} u_D,
\quad
 m_\Dp = \frac{\vHp}{v_H} \times m_D,
\end{align}
so that the Majorana mass dependence only appears
in the logarithmic term from the neutrino loop.
The diagonal matrix $d_n$ and unitary matrix $u_D$ are parametrized as
\begin{align}
 d_n =&\ \mathrm{diag}\left(d_1, d_2, d_3\right), \\
u_D =&\
\begin{pmatrix}
1 & 0 & 0 \\
0 & c^n_{23} & s^n_{23} \\
0 & -s^n_{23} & c^n_{23} \\
\end{pmatrix}
\begin{pmatrix}
c_{13}^n & 0 & s_{13}^n e^{-i \delta_{n}} \\
0 & 1 & 0 \\
-s^n_{13} e^{i \delta_n} & 0 & c^n_{13} \\
\end{pmatrix}
\begin{pmatrix}
 c^n_{12} & s^n_{12}& 0 \\
 -s^n_{12} & c^n_{12} & 0 \\
0 & 0 & 1
\end{pmatrix}.
\end{align}
The parameters in $d_n$ and $u_D$ are fit to
explain the active neutrino masses and mixing.

With these simplifications, the radiative corrections to the neutrino masses are
given by
\begin{align}
 \delta m_{LL}
= &\
\frac{g^2}{64\pi^2 c_W^2} u_D^T d_n
  \left(\frac{m_h^2}{m_Z^2}\log \frac{M_N^2}{m_h^2}-\log\frac{M_N^2}{m_Z^2} \right)
u_D  \\ \notag
 &\ +\frac{g^2}{32\pi^2} V_G D_\chi \frac{\Delta g_{\chi L}}{\Delta m_L^2} V_G^T
 -\frac{gm_W}{16\pi^2 v_u^2} V_G D_\chi
 \left( g_{\chi L} u_D^T d_n u_D + u_D^T d_n u_D g_{\chi L}
 + 2u_D^T d_n u_D L_{\chi Z}\right) V_{H_u}^T,
\end{align}
where
\begin{align}
\Delta = \Delta_0  \times u_D^T d_n u_D, \quad
\Delta_0 := B_0 -2 X_0.
\end{align}
We simply neglect the loop corrections to the mirror neutrinos,
i.e. $\delta m_{\Lp\Lp} = 0$,
since the sterile neutrino masses are dominated by the tree-level contributions
and there are large ambiguities of SUSY breaking parameters in the twin sector.

For concreteness, we consider the CMSSM scenario of the soft parameters with
\begin{align}
 \tan\beta=10,\quad
 \mathrm{sgn}\mu = +1,
 \quad
 m_0 = -A_0 = 5~\TeV,
 \quad
 M_{1/2} = 2.5~\TeV.
\end{align}
We calculate the parameters at TeV scale
by using \texttt{softsusy-4.1.12}~\cite{Allanach:2001kg}.
At this point, the SM-like Higgs mass is $125.69~\GeV$.
The soft parameters relevant to the neutrino masses are given by
\begin{align}
 M_1 = 1.134~\TeV,
\quad
 M_2 = 2.017~\TeV,
\quad
\mu = 3.424~\TeV,
\quad
m_L = (5.2269, 5.2268, 5.1983)~\TeV.
\end{align}
We also studied mirage mediation boundary conditions
to see how these soft parameters change the results,
but the results are not changed significantly among the studied scenarios,
as shown in App.~\ref{sec-mirage}.
The parameters for the right-handed (s)neutrinos are set to
\begin{align}
 M_N = 10^{12}~\GeV,
\quad
\Delta_0 = 3~\TeV.
\end{align}
The parameters in $d_n$ and $u_D$
are fit to explain the neutrino mixing parameters,
i.e. the two mass squared differences and the PMNS matrix.
The lightest neutrino, $\nu_1$ for NO and $\nu_3$ for IO,
mass is assumed to be $0.001~\eV$.
We scan over the value of $v_{H^\prime}/v_H$.
We always found the values
which explain the neutrino mixing parameters throughout our scan,
up to numerical errors.

The left panel of Fig.~\ref{fig-nu4CM} shows
the mass and mixing of the lightest sterile neutrino $\nu_4$.
The solid (dashed) lines are the cases of NO (IO) of the active neutrinos.
Since we assume $m_D \propto m_\Dp$,
the mixing matrix for the sterile neutrinos are similar
to the active ones up to the $\order{1}$ coefficients from the soft parameters,
i.e.
\begin{align}
 \begin{pmatrix}
 \abs{ U_{e4}} \\ \abs{U_{\mu 4}} \\ \abs{U_{\tau 4}}
 \end{pmatrix}
\propto
  \begin{pmatrix}
 \abs{ U_{e1}} \\  \abs{ U_{\mu 1}} \\  \abs{ U_{\tau 1}}
 \end{pmatrix}
~\mathrm{for~NO\quad and}~
 \begin{pmatrix}
 \abs{ U_{e4}} \\ \abs{U_{\mu 4}} \\ \abs{U_{\tau 4}}
 \end{pmatrix}
\propto
  \begin{pmatrix}
 \abs{ U_{e3}} \\  \abs{ U_{\mu 3}} \\  \abs{ U_{\tau 3}}
 \end{pmatrix}
~\mathrm{for~IO}.
\end{align}
Thus, $\abs{U_{e4}}$ ($\abs{U_{\mu 4}}$)
is the largest element in the NO (IO) case.

The right panel of Fig.~\ref{fig-nu4CM} shows the lifetime of $\ndbd$ decays.
Since the sterile neutrinos are much lighter than $\order{100~\MeV}$
for $v_{H^\prime}/v_H \lesssim 10^4$,
the contributions are proportional to $U_{ei}^2 m_{\nu_i}$.
In the NO case, the contributions from the heavier sterile neutrinos
are more suppressed by the mixing angles, see Eq.~\eqref{eq-absPMNS}.
While in the IO case, the heavier states have degenerate masses,
$m_{\nu_5} \simeq m_{\nu_6}$ and the mixing angles are not suppressed.
Therefore the lifetimes are much shorter for the IO case,
and hence $m_{\nu_1} \lesssim 0.001~\eV$ is required to be consistent
with the current limits.

\begin{table}[t]
 \centering
\caption{\label{tab-bench}
The benchmark points in the CMSSM scenario.}
\small
\begin{tabular}[t]{c|cc} \hline
 & NO & IO \\ \hline\hline
$\log_{10} v_{H^\prime}/v_H$ & 0.89 & 0.89 \\
$(d_1, d_2, d_3)~[\mathrm{eV}]$ & (0.0185, -0.1621, 0.9270) & (-0.9129, 0.9271, 0.0185) \\
$(s_{12}^n, s_{23}^n, s_{13}^n, \delta_n)$ & (0.3680, 0.7051, 0.5073, 0.3436) & (0.4601, 0.7575, 0.4231, 0.5122) \\ \hline
$(\Delta m_{12}^2\times 10^5, \Delta m_{23}^2\times 10^{3})~[\mathrm{eV}^2]$ & (7.550, 2.424) & (7.550, -2.500) \\
 $(s_{12}^2, s_{23}^2, s_{13}^2, \delta)$ & (0.320, 0.547, 0.022, -2.478) & (0.320, 0.551, 0.022, -1.379) \\
 $(m_{\nu_4}, m_{\nu_5}, m_{\nu_6})$ [eV] & (1.136, 9.932, 56.784) & (1.136, 55.923, 56.789) \\ \hline
 $\left|\begin{pmatrix} U_{e4} & U_{e5} & U_{e6} \\ U_{\mu4} & U_{\mu5} & U_{\mu6} \\ U_{\tau4} & U_{\tau5} & U_{\tau6} \end{pmatrix}\right|$ & $
 \begin{pmatrix}0.1041 & 0.0714 & 0.0189 \\
0.0401 & 0.0772 & 0.0934 \\
0.0620 & 0.0723 & 0.0850 \\
\end{pmatrix}  $& $
 \begin{pmatrix}0.0191 & 0.1043 & 0.0711 \\
0.0937 & 0.0517 & 0.0696 \\
0.0846 & 0.0524 & 0.0800 \\
\end{pmatrix}$ \\ \hline
$(T^{0\nu}_{1/2}(^{76}\mathrm{Ge}), T^{0\nu}_{1/2}(^{136}\mathrm{Xe}))$ years &$(689.99, 626.67)\times 10^{25}$ & $(3.46, 3.14)\times10^{25}$ \\ \hline
  \end{tabular}
\end{table}

Table~\ref{tab-bench} shows the benchmark points in the NO and IO cases.
The size of $\vHp/v_H$ is chosen
such that the anomaly in the reactor experiments, discussed in the next section,
are explained in the NO case.
We see that the neutrino mixing data is consistent
with the neutrino mixing observables.
The mixing angles involving the sterile neutrinos are much smaller than
those in the active neutrinos, and hence the $3\times 3$ PMNS matrix
is almost unitary.
Since we assume the flavor structure of the Dirac matrices are the same,
the relative sizes of the masses and mixing are similar
among the active and sterile neutrinos.
The lifetime of the eV sterile neutrinos are longer than
$10^{35}~\mathrm{sec}$~\cite{Lee:1977tib,Pal:1981rm,Barger:1995ty,Drewes:2016upu},
so the sterile neutrinos are stable as compared to the age of the universe.

\subsection{Sterile neutrino phenomenology}

\begin{figure}[t]
 \centering
\includegraphics[height=0.6\hsize]{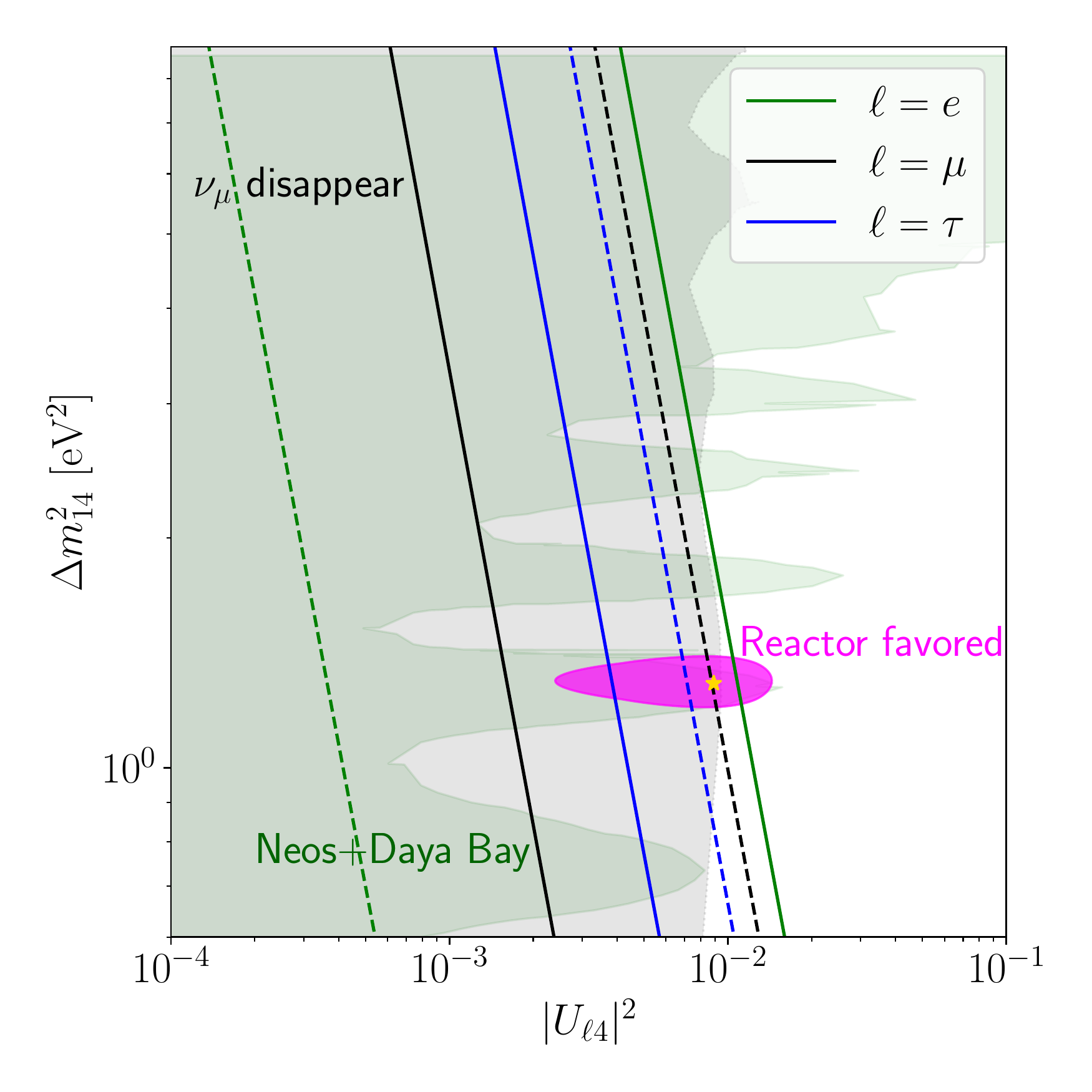}
\caption{\label{fig-nu4obs}
The mixing angle vs $\Delta m_{14}^2 := m_4^2-m_1^2$.
The green, black and blue lines are $\ell = e, \mu$ and $\tau$, respectively.
The solid (dashed) lines are normal (inverse) hierarchy.
The green (gray) region is allowed by the Neos$+$Daya ($\nu_\mu$ disappearance).
The combined $\nu_e$ disappearance result favors the magenta region.
The yellow star is the best-fit point in the analysis of Ref.~\cite{Dentler:2018sju}.
}
\end{figure}

We shall study the phenomenology of the neutrino mixing
with the lightest sterile neutrino $\nu_4$,
see Refs.~\cite{Boser:2019rta,Dasgupta:2021ies}
for recent reviews of sterile neutrinos.
Under the assumption of $m_\Dp \propto m_D$,
the heavier state $\nu_5$ is about 8~(50) times heavier than $\nu_4$
in the NO (IO), and hence the mixing with these will be sub-dominant~\footnote{
See Refs.~\cite{Giunti:2011gz,Hollander:2014iha}
for the analysis with more than two sterile neutrinos.}.
The following combinations of the mixing with $\nu_4$
are constrained from the reactor experiments~\cite{Dentler:2018sju},
\begin{itemize}
 \item {$\nu_e$ disappearance $\propto |U_{e4}|^2$~\cite{DayaBay:2016qvc,NEOS:2016wee,Alekseev:2016llm}}
 \item {$\nu_\mu\to\nu_e$ oscillations at short baseline $\propto |U_{e4} U_{\mu4}|^2$~\cite{LSND:2001aii,MiniBooNE:2013uba,KARMEN:2002zcm,NOMAD:2003mqg,Antonello:2012pq,OPERA:2013wvp}}
 \item {$\nu_\mu$ disappearance, $\propto |U_{\mu4} |^2$~\cite{IceCube:2014flw,IceCube:2016rnb,MINOS:2017cae,NOvA:2017geg,Super-Kamiokande:2010orq,MiniBooNE:2009ozf,MiniBooNE:2012meu,Wendell:2014dka}}
\end{itemize}
The limit for $\abs{U_{\tau 4}}^2 < 0.13$~\cite{Dentler:2018sju}
is much weaker than limits
on the above combinations~\cite{MINOS:2017cae,NOvA:2017geg}.
In the reactor experiments measuring the $\nu_e$ disappearance,
DayaBay and NEOS put upper bounds on $\abs{U_{e4}}^2$~\cite{DayaBay:2016qvc,NEOS:2016wee},
but the DANSS reported an excess~\cite{Alekseev:2016llm}.
It is interesting that the excess can be explained
consistently with the limits from DayaBay+NEOS,
where $\Delta m_{14}^2 = 1.29~\eV^2$ and $\abs{U_{e4}}^2= 0.0089$~\cite{Dentler:2018sju}.
Anomalies are found in the short base-line experiments
LSND~\cite{LSND:2001aii} and MiniBooNE~\cite{MiniBooNE:2013uba},
which favor $\Delta m_{14}^2 \sim 0.5~\eV^2$
and $4|U_{e4} U_{\mu4}|^2 \sim 0.007$.
The $\nu_\mu$ measurements, however, exclude $|U_{\mu 4}|^2 \lesssim 0.01$
for $\Delta m_{14}^2 \sim \order{0.1-10~\eV^2}$,
and therefore the explanation of the short base-line anomalies
by the mixing with a sterile neutrino
is excluded by the $\nu_\mu$ disappearance result~\cite{Dentler:2018sju}.
Hence we do not consider the anomalies in the short base-line experiments.

Figure~\ref{fig-nu4obs} shows the favored regions
by the experiments searching for the mixing with a sterile neutrino.
The green (gray) region is allowed by the Neos+Daya Bay
($\nu_\mu$ disappearance) result,
which should be compared with the green (black) lines.
The pink region is the favored region from all the reactor data,
including the DANSS result which observed the anomaly.
In the NO case, the green solid line overlaps the pink region,
and the black line is inside the gray region.
Thus, the anomaly in the DANSS experiment can be explained in this case.
The benchmark point for the NO case in Table~\ref{tab-bench} is chosen
from the overlapped region.
In the IO case, however, the mixing with electron $\abs{U_{e4}}$
is much smaller than the value preferred by the reactor data
for $\Delta m_{14}^2 \sim 1~\eV$.
Furthermore, this case will be excluded by the $\nu_\mu$ disappearance result
even if $\abs{U_{e4}}$ has a certain value
because $\abs{U_{e4}} < \abs{U_{\mu4}}$.
Therefore, the reactor data is fully explained only in the NO case.

\subsection{Cosmology}

The $\eV$ sterile neutrinos which can explain the reactor anomaly may,
however, be incompatible
with the cosmological observations~\cite{Dasgupta:2021ies}.
The Planck collaboration obtained the $95\%$ C.L. upper limits
on the effective number of neutrinos $N_\eff < 3.29$
and the sum of the neutrino masses $\sum m_\nu < 0.65~\eV$~\cite{Planck:2018vyg}.
These results have strong tension with the reactor anomaly~\cite{Berryman:2019nvr,Adams:2020nue,Hagstotz:2020ukm}.
This tension would be solved by a low reheating temperature $\lesssim 100~\MeV$
after inflation~\cite{Gelmini:2004ah}
or matter domination by some particle, such as moduli fields.
The interactions of the sterile neutrinos
could also resolve the tension~\cite{Dasgupta:2013zpn,Hannestad:2013ana}.
It is also possible that the $\keV$ sterile neutrinos
become the DM due to the mixing with the active neutrinos,
known as the Dodelson-Widrow mechanism~\cite{Dodelson:1993je}.
Since there are three sterile neutrinos in our model,
there would be a case that the reactor anomaly is explained
by one of the sterile neutrinos and the others contribute to the DM.

Before closing, we briefly discuss the cosmology of the other particles in the twin sector.
The twin photon may contribute to $N_\eff$ along with the light sterile neutrinos.
The contribution could be suppressed
if the temperature of the thermal bath of the twin sector
is significantly smaller than the MSSM one.
This require that the reheating process occurs predominantly in the MSSM sector.
Another possibility is that the twin photon is massive
due to the non-zero VEV of the charged Higgs or sparticles.
This would be the case, for instance,
if the anomaly mediation~\cite{Randall:1998uk,Giudice:1998xp}
is the dominant source for the SUSY breaking in the twin sector.
In our model, the photon to twin photon kinetic mixing,
$\eps F^{\mu\nu} F^\prime_{\mu\nu}$,
where $F_{\mu\nu}$ ($F^\prime_{\mu\nu}$) is the field strength of the (twin) photon,
is expected to be tiny.
There is a 3-loop diagram which is mediated by neutrinos
whose order is estimated as
\begin{align}
 \eps \sim \frac{g^3g^{\prime 3} m_{\nu^\prime}^2}{(16\pi^2)^3 m_W m_{W^\prime} }
      \sim 10^{-25}\times
          \left(\frac{ \sqrt{gg^\prime}}{0.5}\right)^6
           \left(\frac{m_{\nu^\prime}}{50~\eV}\right)^2
           \frac{m_{W}}{m_{W^\prime}}.
\end{align}
Therefore it is negligibly small.

The twin electrons and baryons are stable and can contribute to the DM density.
If the asymmetry of particle and anti-particle is negligible,
the twin particles annihilate
when they freeze-out from the twin thermal bath.
While the twin particles become the asymmetric DM
if the asymmetry is non-negligible also in the twin sector.
Thus the abundance of the twin fermions will be small
if the annihilation is large or the asymmetry is small.

The lightest SUSY particle (LSP) in the twin sector
may also be stable due to R-parity in the same way as the LSP in the MSSM.
If $m_{\mathrm{tLSP}} > m_{3/2}$, 
the twin LSP (tLSP) can decay to the gravitino plus the SUSY partner of the tLSP,
or to the MSSM sparticle through the gravitino, depending on the mass spectrum.
For example, the tLSP can decay to a gravitino via the processes
 $\tnu^\prime \to \nu^\prime \psi_{3/2}$ or $\wt{B}^\prime \to \gamma^\prime \psi_{3/2}$
as long as it is kinematically allowed. 
The gravitino can then decay to the LSP in the MSSM.
Here we assume that the $m_{\mathrm{tLSP}} > m_{3/2} > m_\mathrm{LSP}$.
In this case, the twin LSP should be heavier than the TeV scale,
so that the twin LSP/gravitino decay does not alter the success
of Big Bang Nucleosynthesis (BBN).

\section{Conclusion}
\label{sec-concl}

In this paper, we study the neutrino sector
in the model constructed from F-theory.
The model has the MSSM sector and a twin sector,
together with three generations of right-handed neutrinos.
The active neutrino masses are predominantly produced from 1-loop contributions,
because the tree-level active neutrino masses are negligible.
The masses of the sterile neutrinos are generated at the tree-level
with the type-I seesaw mechanism.
We derived the formula for the neutrino mass matrix
at the leading order
in the heavy Majorana mass parameter of the universal right-handed neutrinos.

We analyzed neutrino masses and mixing, as well as the $\ndbd$ decay
under the assumption that the neutrino Yukawa matrix
has a common structure in both the MSSM and twin sectors.
The mixing angles of the lightest sterile neutrino
and the $\ndbd$ decay are shown in Fig.~\ref{fig-nu4CM}.
We showed that the anomaly in the reactor experiments
can be explained only in the NO case,
while the mixing pattern does not match the current experimental data
in the IO case.

The light sterile neutrinos with order $\eV$ mass can explain the anomalies
in the reactor experiments,
but these are disfavored from cosmological observations.
If this is the case, then
the twin EW breaking scale should be so large
that the sterile neutrino masses do not affect the cosmological observables.
Another interesting possibility is that the tension is resolved
by the non-standard history of the universe.
For instance, matter-domination at a late time can dilute the sterile neutrino abundance,
so that these do not affect BBN or the cosmic microwave background.
The study for cosmology,
including the light sterile neutrinos, DM abundance, baryon asymmetry and so on,
is interesting but beyond the scope of this paper.

\section*{Acknowledgment}
This paper is dedicated to the memory of Isaiah F.Johnson.
S.R. greatly appreciates the many discussions with Herb Clemens on the F-theory model.
J.K. thanks to Mohedi Masud for helpful discussions.
The work of J.K.
is supported in part by
the Institute for Basic Science (IBS-R018-D1),
and the Grant-in-Aid for Scientific Research from the
Ministry of Education, Science, Sports and Culture (MEXT), Japan No.\ 18K13534.
The work of S.R. is supported in part by the Department of Energy (DOE) under Award No.\ DE-SC0011726.

\appendix

\section{Results with mirage mediation}
\label{sec-mirage}

To see how the neutrino masses depend on the soft parameters,
we study the two benchmark points
in the so-called mirage mediation~\cite{Choi:2004sx,Choi:2005ge,Endo:2005uy,Choi:2005uz,Falkowski:2005ck}.
In the first case (i), we take
\begin{align}
\tan\beta = 30, \quad
 \mathrm{sgn}\;\mu = +1, \quad
 M_0 = 4.05~\TeV, \quad
 \alpha = 1.5, \quad
c_5 = c_{10} = c_{H_u} = c_{H_d} = 1.0,
\end{align}
where the parameters are defined in Ref.~\cite{Jeong:2021qey}.
The choice of the modular weights are motivated
to enhance the radiative correction
to the SM-like Higgs boson mass~\cite{Abe:2014kla},
which is $125.09~\GeV$ at this point.
The soft parameters at the SUSY breaking scale are
\begin{align}
 M_1 = 4.0348~\TeV,
\quad
 M_2 = 3.8977~\TeV,
\quad
\mu = 0.7869~\TeV,
\quad
m_L = (3.9210,3.92002,3.64118)~\TeV.
\end{align}
In contrast to the CMSSM case,
the Higgsinos are lighter than the gauginos.

The other case (ii) is
\begin{align}
\tan\beta = &\ 30.41, \quad
 \mathrm{sgn}\;\mu = +1, \quad
 M_0 = 1.038~\TeV, \quad
 \alpha = -0.7734, \\ \notag
 c_5 =&\ -0.1951, \quad
c_{10} = 0.06998, \quad
c_{H_u} = 2.892, \quad c_{H_d} = -1.665.
\end{align}
This case is the point (A) in Ref.~\cite{Jeong:2021qey},
and can explain the anomaly
in the muon anomalous magnetic moment~\cite{Muong-2:2021ojo}.
The SM-like Higgs boson mass is $125.1~\GeV$ at this point.
The values of the soft parameters are given by
\begin{align}
 M_1 = 1.612~\TeV,
\quad
 M_2 = 0.741~\TeV,
\quad
\mu = 2073~\TeV,
\quad
m_L = (0.76811,0.76815,0.78143)~\TeV.
\end{align}
This point has relatively light slepton masses.
In this case, $\Delta_0 = 10~\TeV$ to enhance the gaugino loop correction
together with the light sleptons.

Table~\ref{tab-mirage} is the same table as Table~\ref{tab-bench}
in the mirage mediation.
The sterile neutrino masses and mixing
are similar to the CMSSM case in Table~\ref{tab-bench}.
Hence the dependence on the soft parameters are mild
because more than 60\% of the radiative corrections
are originated from the loop corrections enhanced
by the logarithmic factor with the right-handed neutrino mass $\sim L_{NZ}$.
While for the case~(ii), the sterile neutrino masses
are lighter than the CMSSM case,
because the loop corrections are enhanced by larger $\Delta_0$
and light sleptons.
The mixing matrices have similar structure as far as
the soft parameters are flavor universal.

\begin{table}[h]
 \centering
 \caption{\label{tab-mirage}
The benchmark points in the two cases of the mirage mediation.
}
\small
\begin{tabular}[t]{c|cc} \hline
case~(i)  & NO & IO \\ \hline\hline
$\log_{10} v_{H^\prime}/v_H$ & 0.89 & 0.89 \\
$(d_1, d_2, d_3)~[\mathrm{eV}]$ & (0.0187, -0.1635, 0.9353) & (-0.9207, 0.9353, 0.0187) \\
$(s_{12}^n, s_{23}^n, s_{13}^n, \delta_n)$ & (0.3680, 0.7047, 0.5070, 0.3436) & (0.4436, 0.7636, 0.4109, 0.5280) \\ \hline
$(\Delta m_{12}^2\times 10^5, \Delta m_{23}^2\times 10^{3})~[\mathrm{eV}^2]$ & (7.550, 2.424) & (7.550, -2.500) \\
 $(s_{12}^2, s_{23}^2, s_{13}^2, \delta)$ & (0.320, 0.547, 0.022, -2.478) & (0.320, 0.551, 0.022, -1.379) \\
 $(m_{\nu_4}, m_{\nu_5}, m_{\nu_6})$ [eV] & (1.145, 10.018, 57.290) & (1.145, 56.398, 57.291) \\ \hline
 $\left|\begin{pmatrix} U_{e4} & U_{e5} & U_{e6} \\ U_{\mu4} & U_{\mu5} & U_{\mu6} \\ U_{\tau4} & U_{\tau5} & U_{\tau6} \end{pmatrix}\right|$ & $
 \begin{pmatrix}0.1041 & 0.0714 & 0.0189 \\
0.0402 & 0.0773 & 0.0933 \\
0.0620 & 0.0723 & 0.0850 \\
\end{pmatrix}  $& $
 \begin{pmatrix}0.0191 & 0.1057 & 0.0690 \\
0.0937 & 0.0502 & 0.0706 \\
0.0845 & 0.0510 & 0.0809 \\
\end{pmatrix}$ \\ \hline
$(T^{0\nu}_{1/2}(^{76}\mathrm{Ge}), T^{0\nu}_{1/2}(^{136}\mathrm{Xe}))$ years &$(677.86, 615.65)\times 10^{25}$ & $(3.83, 3.48)\times10^{25}$ \\ \hline
\end{tabular}
\\
\vspace{1.0cm}
\begin{tabular}[t]{c|cc}\hline
 & NO & IO \\ \hline\hline
$\log_{10} v_{H^\prime}/v_H$ & 0.89 & 0.89 \\
$(d_1, d_2, d_3)~[\mathrm{eV}]$ & (0.0166, -0.1450, 0.8293) & (-0.8292, 0.8158, 0.0166) \\
$(s_{12}^n, s_{23}^n, s_{13}^n, \delta_n)$ & (0.3680, 0.7041, 0.5064, 0.3437) & (0.8347, 0.5200, 0.7035, -0.4038) \\ \hline
$(\Delta m_{12}^2\times 10^5, \Delta m_{23}^2\times 10^{3})~[\mathrm{eV}^2]$ & (7.550, 2.424) & (7.550, -2.500) \\
 $(s_{12}^2, s_{23}^2, s_{13}^2, \delta)$ & (0.320, 0.547, 0.022, -2.478) & (0.320, 0.551, 0.022, -1.379) \\
 $(m_{\nu_4}, m_{\nu_5}, m_{\nu_6})$ [eV] & (1.015, 8.880, 50.796) & (1.016, 49.972, 50.793) \\ \hline
 $\left|\begin{pmatrix} U_{e4} & U_{e5} & U_{e6} \\ U_{\mu4} & U_{\mu5} & U_{\mu6} \\ U_{\tau4} & U_{\tau5} & U_{\tau6} \end{pmatrix}\right|$ & $
 \begin{pmatrix}0.1041 & 0.0713 & 0.0189 \\
0.0402 & 0.0774 & 0.0933 \\
0.0619 & 0.0723 & 0.0851 \\
\end{pmatrix}  $& $
 \begin{pmatrix}0.0191 & 0.1080 & 0.0654 \\
0.0938 & 0.0479 & 0.0721 \\
0.0845 & 0.0484 & 0.0826 \\
\end{pmatrix}$ \\ \hline
$(T^{0\nu}_{1/2}(^{76}\mathrm{Ge}), T^{0\nu}_{1/2}(^{136}\mathrm{Xe}))$ years &$(862.73, 783.56)\times 10^{25}$ & $(6.05, 5.50)\times10^{25}$ \\ \hline
 \end{tabular}
\end{table}

\clearpage
{\small
\bibliographystyle{JHEP}
\bibliography{refs_twin}
}

\end{document}